\documentclass[a4paper, amsfonts, amssymb, amsmath, reprint, showkeys, nofootinbib, twoside]{revtex4-2}
\usepackage[english]{babel}
\usepackage[utf8]{inputenc}
\usepackage[colorinlistoftodos, color=green!40, prependcaption]{todonotes}
\usepackage[pdftex, pdftitle={Article}, pdfauthor={Author}]{hyperref} % For hyperlinks in the PDF
\usepackage{tabularx}
\begin{document}
\title{Understanding the quantum Rabi ring using analogies to quantum magnetism}

\author{Diego Fallas Padilla}
\email{daf5@rice.edu}
\affiliation{Department of Physics and Astronomy, and Rice Center for
Quantum Materials, Rice University, Houston, Texas 77251-1892, USA} 
\author{Guo-Jing Cheng}
\affiliation{Department of Physics, and Chongqing Key Laboratory for strongly coupled Physics, Chongqing University, Chongqing 401330, China}
\author{Yu-Yu Zhang}
\email{yuyuzh@cqu.edu.cn}
\affiliation{Department of Physics, and Chongqing Key Laboratory for strongly coupled Physics,
Chongqing University, Chongqing 401330, China}
\author{Han Pu}
\email{hpu@rice.edu}
\affiliation{Department of Physics and Astronomy, and Rice Center for
Quantum Materials, Rice University, Houston, Texas 77251-1892, USA}

\begin{abstract}
We map a quantum Rabi ring, consisting of $N$ cavities arranged in a ring geometry, into an effective magnetic model
containing the XY exchange and the Dzyaloshinskii Moriya (DM) interactions.
The analogue of the latter is induced by an artificial magnetic field,
which modulates photon hopping between nearest-neighbor cavities with a phase. 
The mean-field behavior of both systems is almost identical, facilitating the description of the different phases in the quantum optical model through simple arguments of competing magnetic interactions.
For the square geometry ($N=4$) the rich phase diagram exhibits three superradiant phases denoted as ferro-superradiant, antiferro-superradiant and chiral superradiant. In particular, the DM interaction is responsible for the chiral phase in which
the energetically degenerate configurations of the order parameters are similar to the in-plane magnetizations of skyrmions with different helicities.
The antiferro-superradiant phase is suppressed in the triangle geometry ($N=3$) as geometric frustration contributes to stabilize the chiral phase even for small values of the DM interaction. The chiral phases for odd and even $N$ show a different scaling behavior close to the phase transition. The equivalent behavior on both systems opens the possibility of simulating chiral magnetism in a few-body quantum optical platform, as well as understanding one system using the insights gained from the other.
\end{abstract}

\date{\today }
\maketitle

\textit{Introduction --} It is not unusual that two seemingly very different systems are connected by the same underlying physics. Finding such connections can often help us to gain new insights into one system by importing knowledge obtained from the study of the other. Here we show how a light-matter interaction system --- the quantum Rabi ring --- can be mapped to a chiral magnetic system consisting of various kinds of magnetic couplings including, for example, the Dzyaloshinskii-Moriya (DM) interaction~\cite%
{dzyaloshinsky1958,moriya1960} which plays a fundamental role in the study of topological
states in magnetic systems, see for example~\cite{nagaosa2013}. The DM interaction
favors non-collinear spin structures, stabilizing interesting spin textures
such as magnetic skyrmions in chiral magnets~\cite%
{bogdanov1989,bogdanov1994,rossler2006} which have been observed
experimentally, for example, in helimagnets under the presence of an external magnetic
field~\cite{muhlbauer2009,yu2010}. Alongside with the DM interaction, other mechanisms
that can stabilize chiral spin textures are perpendicular easy-axis
anisotropy, geometric frustration, or a combination of all the
aforementioned~\cite{hayami2016,leonov2015}. Although such interesting behaviors
have been observed in magnetic solid state materials, exploring a wide range of 
exchange interaction strengths in a single experiment might not be an easy task,
indicating the relevance of finding other highly controllable experimental platforms 
where magnetic systems could be simulated.

Light-matter interaction systems have been used extensively to explore
many-body quantum phases in different platforms such as cavity and circuit
QED~\cite{Greentree2006,plenio,zhu2020,felicetti}, and cold atoms in optical
lattices~\cite{bloch,cai2021,chen2021}. High tunability and control in these experimental environments makes them attractive candidates to simulate condensed matter magnetic systems. For example, cold atoms in optical lattices have been used to simulate antiferromagnetic spin chains~\cite{simon2011} and frustrated classical magnetism~\cite{struck2011}, while an optical cavity has been used to engineer collective spin exchange interactions~\cite{norcia2018}. 

On top of the already interesting interactions between cavity modes and
atoms, other intriguing many-body phases of quantum systems can be achieved
when the presence of external fields is added to the mix. Recent
experimental advances and theoretical findings have addressed synthesizing
magnetic fields for neutral ultracold atoms~\cite{lin2009,dalibard2011,cao2014} and
photonic systems~\cite{umucalilar2012,wang2016,cai2021,roushan2017,bloch2012}. In particular,
addition of artificial magnetic fields have been proven to unlock the
emergence of exotic phases of matter, such as chiral ground-state currents
of interacting photons in a three-qubit loop~\cite{roushan2017}, chiral phases
in a quantum Rabi triangle \cite{zhang2021}, and fractional quantum Hall
physics in the Jaynes-Cummings Hubbard lattice~\cite{hayward2012,hayward2016,noh2016}. Observation
of these stable chiral phases in optical setups serves as a further motivation to draw connections
between these phenomena and chiral magnetism.

Here we study a quantum Rabi ring and show that it can be mapped into a large-spin magnetic Hamiltonian of Lipkin-Meshkov-Glick (LMG) systems with nearest-neighbor interactions. Through this mapping, the different phases in the quantum Rabi ring can be intuitively understood by studying the competition of the DM interaction and the XY Heisenberg exchange interaction in the equivalent LMG ring. Furthermore, we also show that the behavior of the chiral phases is highly influenced by geometric frustration.

\textit{Quantum Rabi ring --} We consider a system with $N$ cavities
placed in a ring. Each cavity contains a two-level atom and is described by the quantum
Rabi Hamiltonian
\begin{equation}
H_{R,n}=\omega a_{n}^{\dagger }a_{n}+g(a_{n}^{\dagger }+a_{n})\sigma
_{n}^{x}+\frac{\Delta }{2}\sigma _{n}^{z},
\end{equation}%
where $a_{n}$ ($a_{n}^{\dagger }$) is the photon annihilation (creation)
operator of the single-mode cavity with frequency $\omega $ at cavity $n$, $%
g $ is the atom-cavity coupling strength, and $\sigma _{n}^{i}$ are the
Pauli matrices representing the two-level atom at site $n$ with energy
splitting $\Delta $ between levels. The dimensionless coupling strength is defined as $%
g_{1}\equiv g/\sqrt{\Delta \omega }$.

Although quantum phase transitions (QPTs) are often studied in the thermodynamic limit of infinite lattice sites (or infinite number of atoms)~\cite{sachdev2011}, some few-body systems such as the quantum Rabi model~\cite{hwang2015,liu2017,chen2020,braak2011,chen2012} or the Jaynes-Cummings finite lattice systems~\cite{hwang2016} have been proven to undergo photon QPTs in alternative limits such as the the classical oscillator (CO) limit with $\Delta /\omega \rightarrow \infty $~\cite{bakemeier2012}. This is the regime we will focus on in this work.

The quantum Rabi ring Hamiltonian contains photon hopping between the neighboring cavities
\begin{equation}
H_{RR}=\sum_{n=1}^{N}H_{R,n}+\sum_{n=1}^{N}J(e^{i\theta }a_{n}^{\dagger
}a_{n+1}+e^{-i\theta }a_{n+1}^{\dagger }a_{n}),  \label{QRTHAM}
\end{equation}%
where $J$ is the hopping amplitude between nearest-neighbor cavities with a
phase $\theta $, and periodic boundary condition implies $a_{N+1}=a_1$. A thorough description of this system in a triangle ($N=3$) can be found in ~\cite{zhang2021}. An artificial vector potential $A(r)$ leads the photon
hopping terms between nearby cavities $n$ and $m$ to become complex with the
phase given by $\theta =\int_{r_{n}}^{r_{m}}$ $A(r)dr$. The sum of the
phases along the closed loop of the N cavities $N\theta $ is equivalent to
the effective magnetic flux. The artificial magnetic field can be realized by a
periodic modulation of the photon hopping strength dependent on cavities
with tunable resonances~\cite{roushan2017,zhang2021}. The complex phase breaks time-reversal symmetry (TRS) and, as will be shown, is
crucial in leading to the DM interaction in the
mapped magnetic model.

In the classical oscillator limit $\Delta /\omega \rightarrow \infty $, the quantum Rabi model in each cavity
undergoes a quantum phase transition from a normal phase to a superradiant phase~\cite{chen2018,hwang2016,liu2017}. 
For small values of $g_{1}$, the average number of photons in the cavity tends to zero, which is the so-called normal phase. As $g_{1}$ increases to the critical point $%
g_{1c}$, the photon population on each cavity becomes macroscopic (proportional to $\frac{\Delta}{\omega}$), signaling the superradiant phase.
Moreover, the hopping of photons between neighboring cavities
unlocks more exotic superradiant phases with the order parameter
$\langle a_n \rangle$ being site-dependent.

A general description of the mean-field features and excitation spectrum of this model can be done by constructing low-energy effective Hamiltonians for each phase (see Supplemental Material~\cite{[{See Supplemental Material at }][{ for more details.}]supp2022}). After shifting the bosonic operator $%
a_{n}\rightarrow \tilde{a}_{n}+\alpha _{n}$ with the complex mean-field value $\alpha
_{n}=A_{n}+iB_{n}$, the effective low-energy Hamiltonian under the condition $J/\omega \ll 1$ is obtained by projecting to the spin subspace $|\downarrow \rangle $, giving
\begin{eqnarray}\label{effective square}
H_{\text{eff}}^{\downarrow } &=&\sum_{n=1}^{N} \big[ \omega \tilde{a}_{n}^{\dagger }\tilde{a}_{n}-%
\frac{\lambda _{n}^{2}}{\Delta _{n}}\left( \tilde{a}_{n}^{\dagger }+\tilde{a}_{n}\right) ^{2}
\notag \\
&&+J\tilde{a}_{n}^{\dagger }(e^{i\theta }\tilde{a}_{n+1}+e^{-i\theta }\tilde{a}_{n-1}) \big]+E_{g},
\end{eqnarray}%
where $\Delta _{n} \equiv \sqrt{\Delta
^{2}+16g^{2}A_{n}^{2}}$, $\lambda
_{n}\equiv g\Delta /\Delta _{n}$, and the mean-field ground-state energy $E_g$ is given by
\begin{eqnarray}
    E_g &=& \sum_{n=1}^{N}\, \big[\omega (A_n^2+B_n^2) - \frac{1}{2}\sqrt{\Delta^2 + 16g^2A_n^2}  \nonumber \\
    &&+ 2J\cos\theta(A_nA_{n+1}+B_nB_{n+1}) \nonumber \\ && 
+ 2J\sin\theta(B_nA_{n+1}-B_{n+1}A_n) \big] \,.
\end{eqnarray}
Minimization of $E_g$ with respect to $A_n$ and $B_n$ (see Supplemental Material) yields the condition $\omega
B_{n}+J\sin \theta (A_{n+1}-A_{n-1})+J\cos \theta (B_{n+1}+B_{n-1})=0$.
Furthermore,
when $N$ is odd, there are additional conditions $B_{N}=-B_{1}$ and $A_{N}=A_{1}$ that need to be satisfied. This indicates the qualitative difference between even and odd $N$. As we will show later, odd $N$ gives rise to geometric frustration. Finally, we want to stress that,
in the absence of the artificial magnetic field (i.e., $\theta =0$ or $\pm\pi$), $B_n=0$ for all $n$.

\textit{Effective magnetic model: LMG ring --} We now map the effective Hamiltonian~(\ref{effective square}) using the Holstein-Primakoff transformation~\cite{holstein1940} given by $S^z = a^{\dagger}a- S$ and $S_n^{+} = a_n^{\dagger}\sqrt{2S -a_n^{\dagger}a_n}$. In the normal phase $\langle a_n \rangle = A_n + iB_n=0$ and in the classical spin limit $S\rightarrow \infty$, the Holstein-Primakoff transformation can be approximated by $S_n^{+}\approx \sqrt{2S}a_n^{\dagger }$, leading to the magnetic Hamiltonian
\begin{eqnarray}
H_{\mathtt{LMGR}} &=&\sum_{n=1}^{N} \left[\omega S_{n}^{z}-\frac{2g_1^2 \omega}{%
S}(S_{n}^{x})^{2} \right]\nonumber\\
&&+\frac{J}{S}\cos \theta \,\sum_{n=1}^N(S_{n}^{x}S_{n+1}^{x}+S_{n}^{y}S_{n+1}^{y})  \notag \\
&&+\frac{J}{S}\sin \theta \,\sum_{n=1}^N\hat{z}\cdot (%
\vec{S}_{n}\times \vec{S}_{n+1}),  \label{LMGR}
\end{eqnarray}
which is also valid for the region in the superradiance phase not too far away from the normal-superradiance phase boundary.
We denote $H_{\mathtt{LMGR}}$ as the LMG ring model as it is a generalization of the LMG
Hamiltonian~\cite{lipkin1965} with additional nearest-neighbor interactions included. The physical meaning of each term in $H_{\mathtt{LMGR}}$ is quite clear: The two terms in the first line represent an external magnetic along the $z$-axis and the easy-axis anisotropy along the $x$-axis, respectively; the
second line is a typical XY spin exchange interaction which is either ferromagnetic or antiferromagnetic depending on the sign of $%
J\cos \theta $; finally, the last line corresponds to the
DM interaction with the strength $J\sin \theta$. The relative strength between the XY and the DM terms is thus controlled by $\theta$.

\begin{figure}[t]
\includegraphics[scale=0.27]{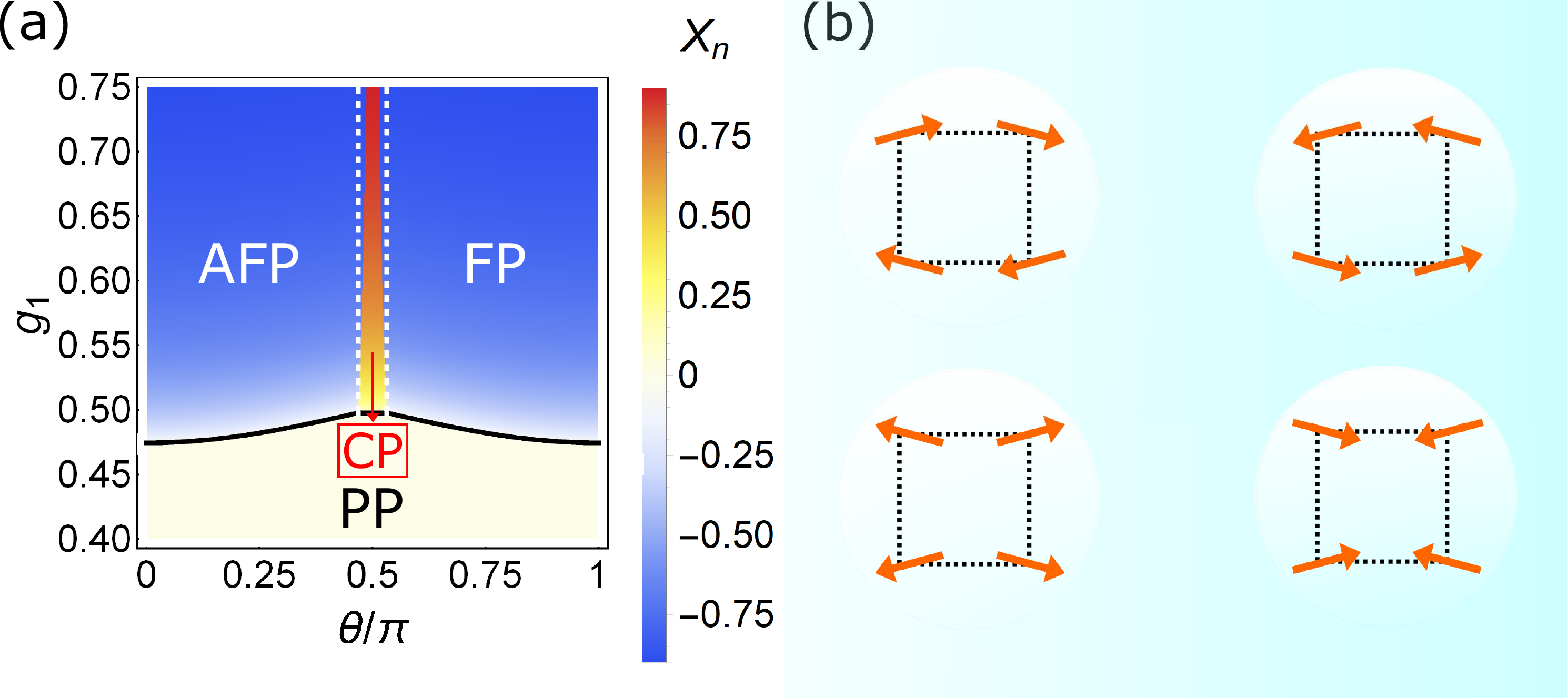}
\caption{(a) Phase diagram in the $\theta$-$g_1$ plane for the LMG ring with $N=4$ using $X_n$ for a given site $n$ as order parameter. To facilitate visualization of the different phases we have chosen one of the degenerate configurations of the ground state for each phase, such that $X_n$ in the chiral phase has opposite sign. The solid black line represents the second-order phase boundary, while vertical dashed lines represent the predicted first-order lines obtained from the equivalent quantum Rabi ring. The hopping strength is set to $J/\omega =0.05$. (b) Allowed magnetization configurations in the CP for $N=4$ represented in the $xy$-plane. }
\label{MFMagnetic}
\end{figure}

Treating $\vec{S}$ as a classical vector, the mean-field energy according to $H_{\mathtt{LMGR}}$ can be readily derived as 
\begin{eqnarray}\label{MFLMGR}
\frac{E_{MF}}{\omega S}&=&\sum_{n=1}^{N} - \sqrt{(1-X_n^2-Y_n^2)} -2g_1^2 X_n^2 \notag \\
&&+\frac{J}{\omega}\cos\theta(X_nX_{n+1}+Y_nY_{n+1}) \notag \\
&&+ \frac{J}{\omega}\sin\theta(X_nY_{n+1}-X_{n+1}Y_n),
\end{eqnarray}
where we have defined $X_n = \frac{\langle S_n^x \rangle}{S}$ and $Y_n = \frac{\langle S_n^y \rangle}{S}$. Minimization $E_{MF}$ with respect to $X_n$ and $Y_n$ yields the ground-state phase diagram. One example with $N=4$ is shown in Fig.~\ref{MFMagnetic}(a). The phase diagram is presented in the $\theta$-$g_1$ parameter space for fixed $J=0.05\omega$. We only consider $\theta \in [0,\pi]$. The phase diagram in the range $\theta \in [-\pi,0]$ is a mirror image of the one presented here. 

For small $g_1$, the system is in the paramagnetic phase (PP) where the spin is polarized by external field term along the $z$-axis. The PP is the analog of the normal phase in the original Rabi ring model. When $g_1$ exceeds a critical value, the system enters various non-paramagnetic phases according to the value of $\theta$ through a second-order phase transition. Defining two critical values of $\theta$ as 
\begin{equation}\label{firstordercrit}
\cos \theta _{c}^{\pm }=\pm \frac{1-\sqrt{1+16J^{2}/\omega ^{2}}}{4J/\omega } \approx \mp 2J/\omega,
\end{equation}
the non-paramagnetic phases can be characterized as:
\begin{enumerate}
    \item Ferromagnetic Phase (FP) --- For $\theta \in (\theta_c^+,\pi]$, and $g_1 > g_{1c}^{\mathtt{F}}=\frac{1}{2}\sqrt{1+\frac{2J}{\omega}\cos \theta}$, the system enters the FP. Here the XY coupling is ferromagnetic. Together with the easy-axis anisotropy term, it polarizes the spin along either the $x$- or the $(-x)$-axis. The ground state in FP is doubly degenerate as a result of the break of the $Z_2$ symmetry.
     \item Antiferromagnetic Phase (AFP) --- For $\theta \in [0, \theta_c^-)$, and $g_1 > g_{1c}^{\mathtt{AF}}=\frac{1}{2}\sqrt{1-\frac{2J}{\omega}\cos \theta}$, the system enters the AFP. Here the XY coupling is antiferromagnetic. The spins are polarized along the $(\pm x)$-axis and neighboring spins are anti-aligned. The ground state in AFP is also doubly degenerate.
     \item Chiral Magnetic Phase (CP) --- In between FP and AFP, for $\theta \in (\theta_c^-,\theta_c^+)$ and $g_1>g_{1c}^{\mathtt{C}}=\frac{1}{2}\sqrt{1+\frac{4J^2}{\omega^2}\sin^2 \theta}$, the DM term dominates over the XY coupling and renders spins at different sites no longer collinear. Here the ground state is 4-fold degenerate, breaking both the $Z_2$ and the $C_4$ symmetries, and the corresponding in-plane magnetization orientation in the $xy$-plane is shown in Fig.~\ref{MFMagnetic}(b). 
\end{enumerate}
Note that the transitions between various non-paramagnetic phases are all of first order, and the critical points at $\theta _{c}^{\pm }$ are triple points where three phases (PP, FP/AFP, and CP) coexist.

 Each magnetic phase described in terms of the values of $X_n$ and $Y_n$, has an equivalent phase in the quantum Rabi ring in terms of $A_n$ and $B_n$, as shown in Table \ref{Equivalent}. A similar phase diagram as in Fig.~\ref{MFMagnetic}(a) would be obtained if we solve the quantum Rabi ring Hamiltonian (\ref{effective square}) directly. In particular, the second-order phase boundary between the normal and the superradiant phases is exactly the same as the boundary between the paramagnetic and non-paramagnetic phases. The first-order phase boundaries between different superradiant phases slightly deviates from those between different non-paramagnetic phases for large $g_1$ far away from the second-order boundary. 

\begin{table}[t]
\caption{Correspondence between phases in the quantum Rabi ring (QRR) and those in the LMG ring (LMGR). }
\begin{tabularx}{0.45\textwidth} { 
  | >{\centering\arraybackslash}X 
  | >{\centering\arraybackslash}X|}
 \hline
 QRR phase& LMGR phase \\
 \hline
 Normal (NP): $A_n=B_n=0$ & Paramagnetic (PP): $X_n=Y_n=0$ \\
 \hline
Ferro-superradiant (FSP): $B_n=0$ and $A_n=A_{n+1}$   & Ferromagnetic (FP): $Y_n=0$ and $X_n=X_{n+1}$   \\
\hline
Antiferro-superradiant (AFSP): $B_n=0$ and $A_n=-A_{n+1}$   & Antiferromagnetic (AFP): $Y_n=0$ and $X_n=-X_{n+1}$  \\
\hline
Chiral superradiant (CSP): $B_n\neq0$ and $A_n \neq 0$   & Chiral magnetic (CP): $Y_n\neq0$ and $X_n\neq0$  \\
\hline
\end{tabularx} 

\label{Equivalent}
\end{table}

The similarity in the mean-field behavior of both systems indicates the possibility of simulating magnetic behavior using the quantum Rabi ring. In addition to realizing various types of magnetic coupling terms, we can also simulate geometric frustration by changing $N$ from even to odd. To this end, let us consider a triangular system with $N=3$. Such an arrangement with antiferromagnetic coupling is a prototypical system that exhibits magnetic frustration.

\begin{figure}[tbp]
\includegraphics[scale=0.27]{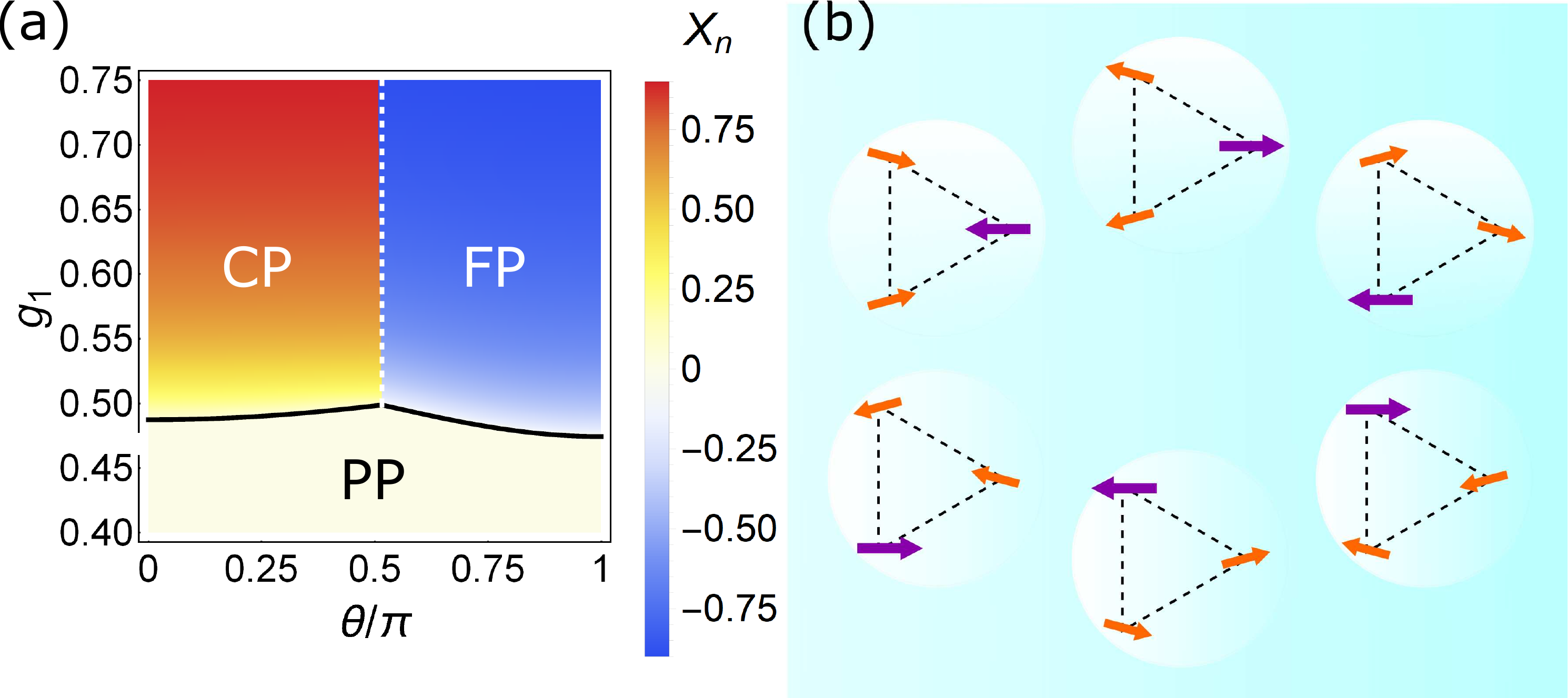}
\caption{(a) Phase diagram in the $\theta$-$g_1$ plane for the LMG ring with $N=3$ using $X_n$ for a given site $n$ as order parameter. To facilitate visualization of the different phases we have chosen one of the degenerate configurations of the ground state for each phase, such that $X_n$ in the chiral phase has opposite sign. The solid black line represents the second-order phase boundary, while vertical dashed lines represent the predicted first-order lines obtained from the equivalent quantum Rabi ring. The hopping strength is set to $J/\omega =0.05$. (b) Allowed magnetization configurations in the CP for $N=3$ represented in the  $xy$-plane. A similar schematic representation can be done for $\alpha_n$ in the complex plane. Arrows with different color and length represent sites with different in-plane magnetization length $l_n=\sqrt{X_n^2+Y_n^2}$.}
\label{QRSPD}
\end{figure}

Following the same procedure, we can obtain the phase diagram for $N=3$ as shown in Fig.~\ref{QRSPD}(a). 
We can use the same conceptual reasoning of competing magnetic interactions to understand the phase diagram of this particular geometry. 
The FP has identical expressions for the second-order phase boundary and order parameter values as the ones found for $N=4$. The region of AFP phase is significantly reduced. In fact, AFP only occurs along the line $\theta=0$. Along this line, the DM interaction vanishes exactly and the spins are aligned in the $x$-direction with the behavior being exactly the one of a classical frustrated anti-ferromagnet described by a Heisenberg model, as pointed out recently~\cite{zhao2022}. 

The CP is much broader in the triangular case as it is defined in the region where $0<\theta \leqslant \theta _{c}=\cos^{-1}\left(-\frac{2J}{\protect\sqrt{8J^2+\protect\omega%
^2}+\protect\omega}\right)$~\cite{zhang2021}. The ground state in the CP is 6-fold degenerate due to the break of both the $Z_2$ and the $C_3$ symmetries. The in-plane magnetization orientation of the degenerate CP states are shown in Fig.~\ref{QRSPD}(b). In the previous square case, the spins at different sites point along different directions, but all have the same length. This is not the case in the triangular case. As one can see from Fig.~\ref{QRSPD}(b), there is always a site where the transverse spin is along the $x$- or $(-x)$-axis and this spin has larger magnitude than the other two. In the triangular quantum Rabi ring model, this means that the photon numbers at different sites are not the same. This has important consequences in the excitation spectrum as we will show later. Also note that,
for small values of $\theta$, the XY exchange interaction $J\cos\theta >0$ is 
stronger than the DM interaction, however, the system still favors a chiral phase with non-collinear
alignment. This is very much in line with observations in antiferromagnetic systems~\cite{hayami2016,mohylna2021,rosales2015} where geometric frustration has been proven to stabilize chiral spin textures favored by the DM interaction, even for very small values of the DM interaction coupling strength.

\begin{figure}[t!]
\includegraphics[scale=0.17]{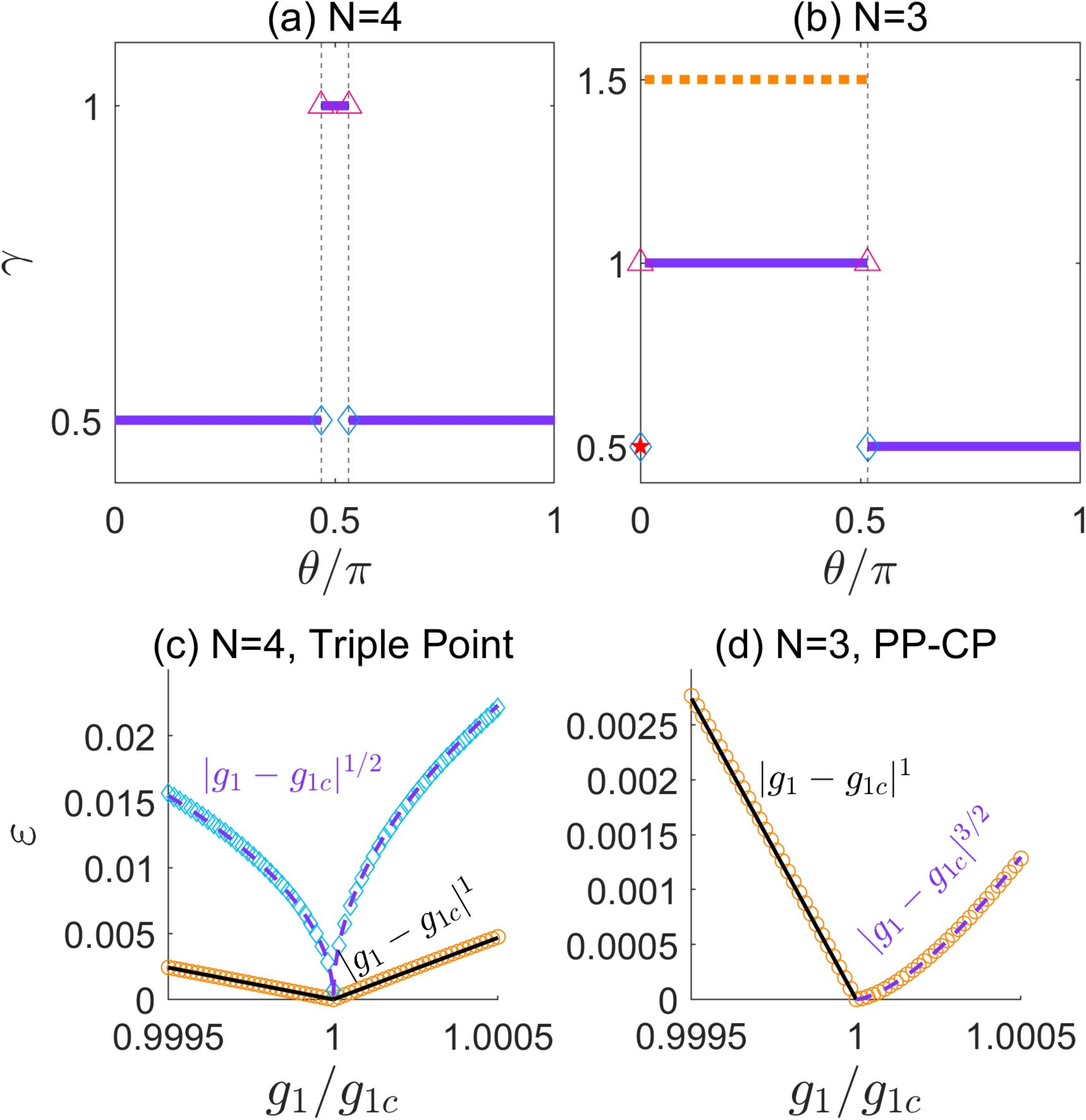}
\caption{Scaling exponents $\gamma$ as a function of $\theta$ for (a) $N=4$ and (b) $N=3$, vertical dashed lines indicate the boundary between non-paramagnetic phases. At the triple points, two modes vanish at the critical point with distinct exponents indicated as open markers. The PP-CP transition for $N=3$ in (b) has non-symmetric exponents, the blue solid and the orange dashed line indicate the exponent when $g_1$ approaches the critical value from below and above, respectively. 
(c) Lowest and second-lowest excitation energies as a function of $g_1/g_{1c}$ across the triple point at $\theta=\theta^+_c$. (d) Lowest excitation energy as a function of $g_1/g_{1c}$ across the PP and the CP boundary for $N=3$. The hopping strength is set to $J/\omega =0.05$, and $\omega=1$.}
\label{excitation energy}
\end{figure}

\textit{Excitation energy scaling --}  To characterize the differences between chiral phase transitions with and without frustration, in particular the consequences of the site-dependent magnetization length $l_n=\sqrt{X_n^2+Y_n^2}$ for $N=3$ shown in Fig.\ref{QRSPD}(b), we analyze the excitation energy behavior near the transition for $N=3$ and $N=4$. The effective Hamiltonian in Eq.~(\ref{effective square}) can be diagonalized as $H_{RSL}=\sum_{n=1}^{N}\varepsilon_n b_{n}^{\dagger }b_{n}$ ~\cite{[{See Supplemental Material at }][{ for more details.}]supp2022}, 
where $\varepsilon_n$ is the excitation energy of the $n$-th mode and $b_n$ ($b_n^{\dagger}$) are bosonic annihilation (creation) operators of such mode obtained through a Bogoliubov transformation. Across a second-order phase boundary we expect the lowest excitation energy $\varepsilon_1$ to vanish exactly at the critical point, with a power-law behavior of the form $\varepsilon_1 \propto |g_1-g_{1c}|^\gamma$ around the critical point $g_{1c}$. Usually, it is expected that the value of $\gamma$ is independent of whether the critical point is approached from above or below. However, as discussed below, this is not the case for the frustrated phases. 

The scaling exponents $\gamma$ as a function of $\theta$ are shown in panels (a) and (b) of Fig.~\ref{excitation energy} for $N=4$ and $N=3$, respectively. For $N=4$ the scaling exponents before and after the transition are always equal to each other. The PP-FP and PP-AFP transitions have the same scaling exponent $\gamma=1/2$, which is the same as the conventional single-cavity Dicke transition \cite{Emary03}. By contrast, $\gamma=1$ for the PP-CP transition. At the triple point, two modes should vanish at the critical point as a signature of the coexistence of both non-paramagnetic phases, see Fig.~\ref{excitation energy}(c). These modes possess exponents 1 and 1/2, respectively, as indicated by the open markers in panel (a).

For $N=3$ in the PP-FP transition $\gamma=1/2$ just as in the $N=4$ case. However, the PP-CP has an unusual scaling behavior as a consequence of the frustrated ground state configurations. As shown in panel (b) and in more detail in panel (d), the exponents at the two sides of the phase transition are different: $\gamma=1$ ($=1.5$) for $g_1$ approaching $g_{1c}$ from below (above). At the triple point, we again have two modes vanishing at the critical point, each of which has a well defined scaling exponent, given by $\gamma=1$ and $\gamma=1/2$, just as for $N=4$.

The difference in the scaling behavior of the CSP for $N=3$ and $N=4$ can be understood by exploring Eq.\ref{effective square}, used to obtained the diagonalized Hamiltonian. For $N=3$ the Hamiltonian in Eq.\ref{effective square}, is $C_3$ symmetric for $g_1<g_{1c}$ as $\lambda_n/\Delta_n$ is independent of $n$, however, for $g_1>g_{1c}$ one of the sites has a different value of $\lambda_n/\Delta_n$, breaking the symmetry. For $N=4$, on the other hand, $A_n^2$ is independent of $n$ in all phases, and consequently,  Eq.~(\ref{effective square}) is always $C_4$ symmetric for both normal/paramagnetic and superradiant/non-paramagnetic phases.

This non-symmetric scaling behavior as a consequence of frustration has been reported for the special point $\theta=0$ \cite{zhao2022} where two modes vanish at the critical point, one with a single exponent $1/2$ and the other with non-symmetric exponents $\gamma=1/2$ and $\gamma=1$ below and above the transition, these three exponents are signaled with open markers in panel (b). This is consistent with our results as the point $\theta=0$ in our model where $\theta$ is a variable represents a triple point between the frustrated AFP and the frustrated CP, then, the vanishing of two modes is expected. Moreover, the non-symmetric $\gamma$ values (1/2 and 1) at this point are different from the ones inside the CP region (1 and 1.5) as the ground-state configurations at $\theta=0$ are not chiral, even though they are still frustrated.

\textit{Conclusions --} We have explored the connection between the
quantum Rabi ring model and a large-spin magnetic model (LMG ring) containing the XY Heisenberg exchange interactions
and the DM interactions.
The mapping between the two systems facilitates a deeper understanding of the phases in
the light-atom coupling system through simple arguments of competing magnetic
interactions, allowing us to connect each phase in the quantum Rabi ring to an equivalent magnetic phase in the LMG ring (Table~\ref{Equivalent}). For a square lattice ($N=4)$, in the region where the DM interaction dominates, a chiral superradiant phase (CSP) is observed. In this phase
$\alpha_n=A_n+ iB_n$ has an orientation that follow typical in-plane magnetization 
patterns observed in skyrmions with different helicities, when represented in the complex plane. Although both $N=4$ and $N=3$ possess a CSP, we find that the phase transitions leading to them are different from each other as indicated by the associated scaling exponents before and after the transition. The presence of geometric frustration for $N=3$ is the key in this distinctive behavior as it stabilizes the chiral phase for a broader region and causes a site-dependent mean-field photon population $|\alpha_n|^2$. This work opens new possibilities on
simulation of magnetic systems using quantum optical platforms. In
particular, the classical oscillator limit considered here, facilitates the
study of systems consisting of only a few (small $N$) spins, which can be a
powerful tool for identifying the building blocks of more complex behaviors
in real materials.

H.P.
acknowledges support from the US NSF and the Welch Foundation (Grant No. C-1669). Y.Y.Z. ackhowledges support from NSFC under Grant No.12075040 and No. 12147102,
Chongqing NSF under Grants No. cstc2020jcyj-msxmX0890, and Fundamental
Research Funds for the Central Universities Grant No. 2021CDJQY-007.

\bibliography{refs}{}

\newpage
\onecolumngrid
\section{Supplemental material: Understanding the quantum Rabi ring using analogies to quantum magnetism}

In this Supplementary Material, we provide useful details about the
derivation of the effective Hamiltonians for the quantum Rabi ring, the mapping between the
LMG ring and the quantum Rabi ring, the derivation of the phase boundaries' expressions, a numerical description of the LMG ring phase diagram,
and the derivation of the analytical values of the mean-field order parameters of the quantum Rabi ring
for the case $N=4$.

\subsection{Effective Hamiltonians of the quantum Rabi ring}
In this section, we derive the effective Hamiltonian of the quantum Rabi ring
in the super-radiant phase, given in Eq.($3$) of the main text.

As the scaled coupling strength $g_{1}=\frac{g}{\sqrt{\Delta \omega}}$ increases to the critical point $%
g_{1c}$, photons in each cavity are macroscropically populated. 
Due to the macroscopic population, the bosonic operator $a_{n}^{\dagger }$ $%
\left( a_{n}\right) $ is expected to be shifted as
\begin{equation}
a_{n}^{\dagger }\rightarrow \tilde{a}_{n}^{\dagger }+\alpha _{n}^{\ast
},\;\;a_{n}\rightarrow \tilde{a}_{n}+\alpha _{n},
\end{equation}%
with the complex displacement $\alpha _{n}=A_{n}+iB_{n}$. One obtains the
Hamiltonian $H_{\text{RR}}=H_{s}+H_{1}$, which consists of the purely atomic part
of the Hamiltonian
\begin{equation}
H_{s}=\sum_{n=1}^N g\left( \alpha _{n}^{\ast }+\alpha _{n}\right) \sigma _{n}^{x}+%
\frac{\Delta }{2}\sigma _{n}^{z},
\end{equation}%
and the remaining Hamiltonian%
\begin{eqnarray}
H_{1} &=&\sum_{n=1}^N\omega (\tilde{a}_{n}^{\dagger }+\alpha _{n}^{\ast })(\tilde{a}_{n}+\alpha
_{n})+\sum_{n=1}^Ng\left( \tilde{a}_{n}^{\dagger }+\tilde{a}_{n}\right) \sigma _{n}^{x}  \notag
\\
&&+J\sum_{n=1}^N[(\tilde{a}_{n}^{\dagger }+\alpha _{n}^{\ast })[e^{i\theta
}(\tilde{a}_{n+1}+\alpha _{n+1})+e^{-i\theta }(\tilde{a}_{n-1}+\alpha _{n-1})].
\end{eqnarray}%
To diagonalize $H_{s}$, we apply the transformation
\begin{equation}
|+z\rangle =\cos \gamma _{n}|e\rangle +\sin \gamma _{n}|g\rangle
,\;\;|-z\rangle =-\sin \gamma _{n}|e\rangle +\cos \gamma _{n}|g\rangle ,
\end{equation}%
with $\tan 2\gamma _{n}=4gA_{n}/\Delta $. This leads to the renormalized
frequency of the $n$-th atom
\begin{equation}
\Delta _{n}=\sqrt{\Delta ^{2}+16g^{2}A_{n}^{2}},
\end{equation}%
and the Pauli matrices can be expressed in the subspace of $\{|+z\rangle
,|-z\rangle \}$ as $\tau _{z}^{n}=\Delta /\Delta _{n}\sigma
_{n}^{z}+4gA_{n}/\Delta _{n}\sigma _{n}^{x}$. Thus the Hamiltonian $H_{RR}$ becomes
\begin{equation*}
H_{\text{RR}}=\sum_{n=1}^{N}\omega \tilde{a}_{n}^{\dagger }\tilde{a}_{n}+\frac{\Delta _{n}}{%
2}\tau _{n}^{z}+\lambda _{n}\left( \tilde{a}_{n}^{\dagger }+\tilde{a}_{n}\right) \tau
_{n}^{x}+J\tilde{a}_{n}^{\dagger }(e^{i\theta }\tilde{a}_{n+1}+e^{-i\theta
}\tilde{a}_{n-1})+V_{off}^{\prime }+E_{0},
\end{equation*}%
where the effective coupling strength is $\lambda _{n}=g\Delta /\Delta _{n}$%
. The off-diagonal term is
\begin{equation}\label{offdiagonal}
V_{off}=\sum_{n=1}^{N}\omega (\alpha _{n}\tilde{a}_{n}^{\dagger }+\alpha _{n}^{\ast
}\tilde{a}_{n})+g\left( \tilde{a}_{n}^{\dagger }+\tilde{a}_{n}\right) \sin (2\gamma _{n})\sigma
_{n}^{z}+J[\tilde{a}_{n}^{\dagger }(e^{i\theta }\alpha _{n+1}+e^{-i\theta }\alpha
_{n-1})+h.c.],
\end{equation}%
and $E_{0}=\sum_{n=1}^{N}\omega \alpha
_{n}^{\ast }\alpha _{n}+J\alpha _{n}^{\ast }(e^{i\theta }\alpha
_{n+1}+e^{-i\theta }\alpha _{n-1})$.

By making off-diagonal term vanish (see below), the transformed Hamiltonian
becomes
\begin{equation}
H_{\text{RR}}=\sum_{n=1}^NH_{R,n}+J\tilde{a}_{n}^{\dagger }(e^{i\theta
}\tilde{a}_{n+1}+e^{-i\theta }\tilde{a}_{n-1})+E_{0},
\end{equation}%
where the transformed quantum Rabi Hamiltonian is%
\begin{equation}
H_{R,n}=\omega \tilde{a}_{n}^{\dagger }\tilde{a}_{n}+\frac{\Delta _{n}}{2}\tau
_{n}^{z}+\lambda _{n}\left( \tilde{a}_{n}^{\dagger }+\tilde{a}{n}\right) \tau _{n}^{x}.
\end{equation}

We perform a Schrieffer-Woff transformation with the unitary operator $%
S_{n}=\exp [-i\sigma _{n}^{y}\lambda _{n}/\Delta _{n}\left( \tilde{a}_{n}^{\dagger
}+\tilde{a}_{n}\right) ]$ on the Rabi Hamiltonian $H_{R,n}$ to eliminate the
block-off-diagonal interaction $\lambda _{n}\left( a_{n}^{\dagger
}+a_{n}\right) \tau _{n}^{x}$. In each cavity, the quantum Rabi Hamiltonian is
transformed as $H_{R,n}^{\prime }=S_{n}^{\dagger
}H_{R,n}S_{n}$. By performing the second-order correction, the transformed
Hamiltonian becomes
\begin{equation}
H_{R,n}^{\prime }=\omega \tilde{a}_{n}^{\dagger }\tilde{a}_{n}+\frac{\Delta _{n}}{2}\tau
_{n}^{z}+\frac{\lambda _{n}^{2}}{\Delta _{n}}(\tilde{a}_{n}+\tilde{a}_{n}^{\dagger
})^{2}\sigma _{n}^{z}+O\left(\frac{\lambda _{n}^{4}}{\Delta _{n}^{4}}\right).
\end{equation}%
Using the unitary transformation $U=\prod_{n}^{N}S_{n}$, we obtain the
effective quantum Rabi ring Hamiltonian
\begin{equation}\label{TH}
H_{\text{RR}}=\sum_{n=1}^{N}H_{R,n}^{\prime }+J\sum_{n=1}^{N}(e^{i\theta }a_{n}^{\dagger }a_{n^{\prime }}+h.c.)+E_{0},
\end{equation}%
where higher-order terms of the second term are neglected for small value of $%
J/\omega $ in the $\Delta /\omega \rightarrow \infty $ limit. 

Since the transformed Hamiltonian (~\ref{TH}) is free of coupling terms between spin states $%
|\downarrow \rangle $ and $|\uparrow \rangle $, the lower-energy Hamiltonian
is obtained by projecting to the spin subspace $|\downarrow \rangle $,
giving
\begin{equation}\label{SEFFSR}
H_{\text{eff}}^{\downarrow }=\sum_{n=1}^{N}\omega \tilde{a}_{n}^{\dagger }\tilde{a}_{n}-%
\frac{\lambda _{n}^{2}}{\Delta _{n}}\left( \tilde{a}_{n}^{\dagger }+\tilde{a}_{n}\right)
^{2}+J\tilde{a}_{n}^{\dagger }(e^{i\theta }\tilde{a}_{n+1}+e^{-i\theta }\tilde{a}_{n-1})+E_{g},
\end{equation}%
The above Hamiltonian is Eq.($3$) of the main text. The ground-state energy
is
\begin{eqnarray}\label{SMFQRR}
E_{g} &=&\sum_{n=1}^{N}\omega (A_{n}^{2}+B_{n}^{2})-\frac{1}{2}\sqrt{\Delta
^{2}+16g^{2}A_{n}^{2}}  \notag \\
&&+2J[(A_{n}A_{n+1}+B_{n}B_{n+1})\cos \theta +\sin \theta
(B_{n}A_{n+1}-B_{n+1}A_{n})]
\end{eqnarray}

Which is Eq.($4$) of the main text.

\subsection{Effective magnetic model }
In this section, we explain the mapping between the paramagnetic phase in the LMG ring and the normal phase in the quantum Rabi ring. We start by noting that in the normal phase $\alpha_n=0$, and, consequently, $a_n =\tilde{a}_n$. The effective Hamiltonian for the quantum Rabi ring in this phase is simply

\begin{equation}
H_{\text{eff}}^{\downarrow }=\sum_{n=1}^{N}\omega a_{n}^{\dagger }a_{n}-
g_1^2 \omega \left( a_{n}^{\dagger }+a_{n}\right)^{2}+Ja_{n}^{\dagger }(e^{i\theta }a_{n+1}+e^{-i\theta }a_{n-1}),
\label{SNPEffective}
\end{equation}

The Lipkin-Meshkov-Glick (LMG) model was proposed initially to describe phase transitions in nuclei~\cite{lipkin1965}, and its Hamiltonian is given by
\begin{equation}
 H_{\text{LMG}} = -\frac{1}{2S}[\gamma_x (S^x)^2 + \gamma_y (S^y)^2] + h S^z
\end{equation}%
where $S^i$ are the components of a spin operator of length $S$. If $S^i$ are regarded as collective spins comprised obtained as a sum of individual $1/2$ spins, the first term of the LMG model describes an infinite range interaction between individual spins, whose anisotropy is modulated by parameters $\gamma_x$ and $\gamma_y$. $h$ represents a transverse field interaction.

Now, let's consider that we have a ring with each site labeled by $n=1,2,..,N$, and each site contains a system described by an LMG Hamiltonian. If interactions between nearest-neighboring LMG systems are allowed, the full Hamiltonian is
\begin{eqnarray}
 H_{\text{LMGR}} &=& \sum_{n=1}^N -\frac{\gamma h}{2S}(S_n^x)^2 + h S_n^z + \frac{\mathcal{J}}{2S}\sum_{n=1}^N(S_n^xS_{n+1}^x+S_n^yS_{n+1}^y) \notag \\
 && +\frac{1}{2S}\sum_{n=1}^N \vec{\mathcal{D}}\cdot(\vec{S}_n \times \vec{S}_{n+1})
 \label{SLMGR}
\end{eqnarray}%
where we have set $\gamma_x=\gamma h$ and $\gamma_y=0$. The third term in Eq.~\ref{SLMGR} is a conventional XX symmetric spin exchange interaction which can be regarded as ferromagnetic or antiferromagnetic depending on the sign of $\mathcal{J}$. Additionally, the last term corresponds to the Dzyaloshinskii-Moriya interaction. From now on, we define the vector $\vec{\mathcal{D}}$ to be constant and pointing along the $z$-direction, namely, $\vec{\mathcal{D}}=(0,0,\mathcal{D})$. 

Here, we are interested in the classical spin limit $S\rightarrow \infty$ mean-field behavior of the model on Eq.~\ref{SLMGR}. In the paramagnetic phase, $\langle S_n^x \rangle = \langle S_n^y \rangle=0$, the Holstein-Primakoff transformation~\cite{holstein1940} can be approximated by

\begin{eqnarray}
S_n^z = a^{\dagger}_n a_n - S, \quad S_n^x \approx \sqrt{\frac{S}{2}}(a_n + a^{\dagger}_n).
 \label{SHP}
\end{eqnarray}%

Using this form of the Holstein-Primakoff transformation on Eq.~\ref{SLMGR}, one obtains the Hamiltonian

\begin{equation}
H_{\text{eff}}=\sum_{n=1}^{N}h a_{n}^{\dagger }a_{n}-
\frac{\gamma h}{2}\omega \left( a_{n}^{\dagger }+a_{n}\right)^{2}+ a_{n}^{\dagger }\left(\left(\frac{\mathcal{J}}{2}+i\frac{\mathcal{D}}{2}\right)a_{n+1}+\left(\frac{\mathcal{J}}{2}-i\frac{\mathcal{D}}{2}\right)a_{n-1}\right),
\end{equation}\label{SNPEFF}

which is exactly the same effective Hamiltonian as the one in Eq.~\ref{SNPEffective} once the parameters are defined as $h=\omega$, $\frac{\gamma}{2}=2g_1^2$, $\frac{\mathcal{J}}{2}=J\cos\theta$, and $\frac{\mathcal{D}}{2}=J\sin\theta$. With these parameter values the Hamiltonian on Eq.~\ref{SLMGR} becomes Eq.($5$) of the main text. The mapping makes evident that the paramagnetic phase in the LMG ring and the normal phase in the quantum Rabi ring are described by the same effective Hamiltonian.

The classical spin limit mean-field energy of the LMG ring with these parameter values is given by

\begin{eqnarray}\label{SMFLMGR}
\frac{E_{MF}}{\omega S}&=&\sum_{n=1}^{N} - \sqrt{(1-X_n^2-Y_n^2)} -2g_1^2 X_n^2
\notag \\
&&+\frac{J}{\omega}\cos\theta(X_nX_{n+1}+Y_nY_{n+1}) \notag \\ 
&&+ \frac{J}{\omega}\sin\theta(X_nY_{n+1}-X_{n+1}Y_n),
\end{eqnarray}%

which is Eq.($6$) of the main text. Here, $X_n = \langle S_n^x \rangle/S$ and $Y_n = \langle S_n^y \rangle/S$.

\subsection{Analytical expressions for the phase boundaries}

The starting point is the effective Hamiltonian on Eq.~\ref{SNPEffective} used to describe the normal phase in the quantum Rabi ring and the paramagnetic phase in the LMG ring. We introduce a Fourier transform defined by $a_n^{\dagger}=\frac{1}{\sqrt{N}}\sum_q e^{inq}a_q^{\dagger}$, the effective Hamiltonian is transformed into

\begin{equation}
H_{\text{eff}}=\sum_{q} \omega_q a_q^{\dagger}a_q - g_1^2 \omega (a_q a_{-q} + a^{\dagger}_q a^{\dagger}_{-q}) - g_1^2\omega,
\end{equation}

 where $\omega_q=\omega -2g_1^2\omega +2J\cos(\theta-q)$. In order to eliminate the non-diagonal terms we can perform a two-mode squeezing transformation given by $S_q = \exp(\xi_q a^{\dagger}_q a^{\dagger}_{-q} -\xi_q^* a_q a_{-q})$, where $\xi_q = r_q e^{i \phi_q}$ being the squeezing parameter. Setting $\phi_q=0$ and $r_q = \frac{1}{8}\log\left(\frac{\omega_q + \omega_{-q}+4g_1^2 \omega}{\omega_q - \omega_{-q}+4g_1^2 \omega}\right)$, the Hamiltonian takes the diagonal form
 
 \begin{equation}\label{diagonalQ}
H_{\text{eff}}=\sum_{q} \epsilon_q a_q^{\dagger}a_q + E_q,
\end{equation}

where $E_q = -g_1^2\omega+\frac{1}{2}\sum_q(\epsilon_q - \omega_q)$, and the excitation energy $\epsilon_q$ is given by

 \begin{equation}
\epsilon_q = \frac{1}{2}\left(\omega_q - \omega_{-q} + \sqrt{(\omega_q + \omega_{-q})^2 - 16g_1^4\omega^2 }\right).
\end{equation}

The specific values that $q$ can take, are determined by the size of the ring $N$. Each superradiant/non-paramagnetic phase is identified by one or more values of $q$. A second-order phase boundary is found when the gap between the ground state and first excited state vanishes. In the case of a Hamiltonian of the form of Eq.~\ref{diagonalQ} the second order boundary is determined by $\epsilon_q=0$, which for a specific $q$ gives the boundary equation

\begin{equation}\label{Scriticalg}
g_{1c}(q,\theta) = \frac{1}{2}\sqrt{\frac{1+\frac{4J}{\omega}\cos\theta\cos q + \frac{4J^2}{\omega^2}\cos(\theta+q)\cos(\theta-q)}{1+2J/\omega \cos\theta \cos q}}.
\end{equation}

This value of $g_1$ signals the second-order boundary between the superradiant/non-paramagnetic phase, characterized by the value $q$, and the normal/paramagnetic phase. Now, if there are two superradiant/non-paramagnetic phases separated by a phase boundary, and such phases are characterized by $q$ and $q'$, the value of $\theta$ where the triple point separating the three phases occurs is given by

\begin{equation} \label{STC}
g_{1c}(q,\theta) = g_{1c}(q',\theta).
\end{equation}

 In the last section of the Supplementary Material and in the main text we present specific values of $\theta_c$ and $g_{1c}$ for $N=4$. 

Now, for the superradiant phases we can conduct a similar procedure as the one just described. Our starting point is now the general effective Hamiltonian on Eq.\ref{SEFFSR}. Now, it might seem that doing the Fourier transform is not possible as $\lambda_n$ and $\Delta_n$ depend on $n$, however, for all the phases in the square lattice (this doesn't follow for $N=3$) the value of $A_n$ depends on $n$ but the value of $A_n^2$ is independent of $n$, and since $\lambda_n$ and $\Delta_n$ depend only on $A_n^2$ turns out both $\lambda_n$ and $\Delta_n$ are independent of n. 

Then, by setting $\lambda_n=\lambda'$ and $\Delta_n=\Delta'$, one can obtain the diagonalized Hamiltonian

 \begin{equation}\label{diagonalQ2}
H_{\text{eff}}=\sum_{q} \epsilon_q' a_q^{\dagger}a_q + E_q',
\end{equation}

where $E_q' = -\frac{\lambda^{'2}}{\Delta'}+\frac{1}{2}\sum_q(\epsilon_q - \omega'_q)$, $\omega_q'=\omega - 2\frac{\lambda^{'2}}{\Delta'} + 2J\cos(\theta-q)$, and the excitation energy $\epsilon'_q$ is given by

 \begin{equation}
\epsilon'_q = \frac{1}{2}\left(\omega'_q - \omega'_{-q} + \sqrt{(\omega'_q + \omega'_{-q})^2 - 16\left(\frac{\lambda^{'2}}{\Delta'}\right)^2 }\right).
\end{equation}

\begin{figure}[t]
\includegraphics[scale=0.4]{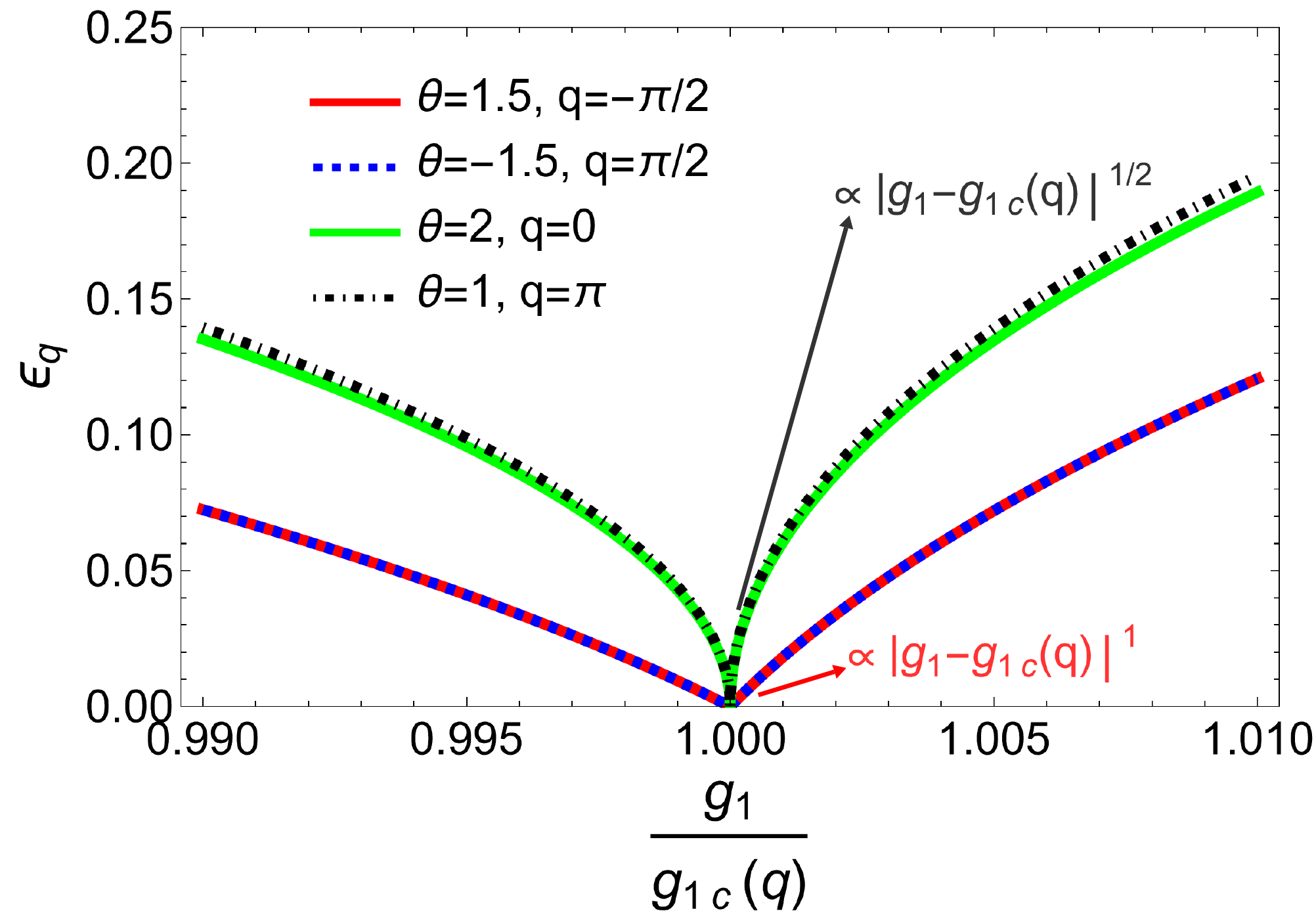}
\caption{Lowest excitation energy for the different phases of the quantum Rabi square across the second order boundary $g_{1c}(q)$, with $q=0$ for the FSP, $q=\pi$ for the AFSP and $q=\pm \pi/2$ for the CSP. The scaling exponents of each phase are indicated with arrows, being $1/2$ for the AFSP and FSP, and $1$ for the CSP.}
\label{Exci}
\end{figure}

Since now we have expressions for $\epsilon_q$ before and after the phase transition boundary $g_{1c}$, we can plot the behavior around $g_{1c}$ for each value of $q$ and study the scaling exponents of $\epsilon_q$ if we represent it as $\epsilon_q \propto |g_1 - g_{1c}(q)|^{\gamma_q}$. For a square lattice $N=4$ the allowed values of $q$ are $0$, $\pm \pi/2$ and $\pi$. The FSP is associated with $q=0$, the AFSP with $q=\pi$ and the CSP with $q=\pm \pi/2$. The behavior of $\epsilon_q$ around $g_{1c}$ for four fixed values of $\theta$ characteristic of each phase is shown in Fig.\ref{Exci}.

As shown in the figure $\gamma_q =1/2$ for $q=0,\pi$ and $\gamma_q=1$ for $q=\pm \pi/2$, consistent with the numerical results presented in Fig.(3) of the main text. Note that the exponents agree with the ones in real space presented in Fig.(3) of the main text. The reason why real space diagonalization is needed is because the Hamiltonian \ref{SEFFSR} for the chiral-phase in the triangle case cannot be Fourier transformed in a similar way as the $\lambda_n^2/\Delta_n$ coefficient is $n$ dependent. The diagonalization in real space, performed to obtain Fig.(3) of the main text, is done as follows. 

The effective Hamiltonian  for $g_{1}>g_{1c}$ \ in Eq. (3) of the main
text bilinear in the creation and annihilation operators $a_{n}^{\dagger }$
and $a_{n}$. It can be diagonalized by the bosonic Bogoliubov
transformation. Denoting $\alpha
=\{a_{1},a_{2},a_{3},a_{4},a_{1}^{\dagger },a_{2}^{\dagger },a_{3}^{\dagger
},a_{4}^{\dagger }\}$, the Hamiltonian in Eq. (3) of the main text for $N=4$
can be written in matrix form as $H_{\text{cCP}}^{\downarrow }=\alpha
M\alpha ^{\dagger }-2(\omega -\lambda _{n}^{\prime 2})$ with $\lambda
_{n}^{\prime 2}=\lambda_{n}^{ 2}/\Delta _{n}$, where the
matrix $M$ \ is given by
\begin{equation}
M(N=4)=\left(
\begin{array}{cccccccc}
\omega /2-\lambda _{1}^{\prime 2} & Je^{-i\theta }/2 & 0 & Je^{i\theta }/2 &
-\lambda _{1}^{\prime 2} & 0 & 0 & 0 \\
Je^{i\theta }/2 & \omega /2-\lambda _{2}^{\prime 2} & Je^{-i\theta }/2 & 0 &
0 & -\lambda _{2}^{\prime 2} & 0 & 0 \\
0 & Je^{i\theta }/2 & \omega /2-\lambda _{3}^{\prime 2} & Je^{-i\theta }/2 &
0 & 0 & -\lambda _{3}^{\prime 2} & 0 \\
Je^{-i\theta }/2 & 0 & Je^{i\theta }/2 & \omega /2-\lambda _{4}^{\prime 2} &
0 & 0 & 0 & -\lambda _{4}^{\prime 2} \\
-\lambda _{1}^{\prime 2} & 0 & 0 & 0 & \omega /2-\lambda _{1}^{\prime 2} &
Je^{i\theta }/2 & 0 & Je^{-i\theta }/2 \\
0 & -\lambda _{2}^{\prime 2} & 0 & 0 & Je^{-i\theta }/2 & \omega /2-\lambda
_{2}^{\prime 2} & Je^{i\theta }/2 & 0 \\
0 & 0 & -\lambda _{3}^{\prime 2} & 0 & 0 & Je^{-i\theta }/2 & \omega
/2-\lambda _{3}^{\prime 2} & Je^{i\theta }/2 \\
0 & 0 & 0 & -\lambda _{4}^{\prime 2} & Je^{i\theta }/2 & 0 & Je^{-i\theta }/2
& \omega /2-\lambda _{4}^{\prime 2}%
\end{array}%
\right) .  \label{Mmatrix}
\end{equation}%
We perform a Bogoliubov's transformation to give bosonic operators $\beta
=\{b_{1}^{\dagger },b_{2}^{\dagger },b_{3}^{\dagger },b_{4}^{\dagger
},b_{1},b_{2},b_{3},b_{4}\}$ as a linear combination of $\alpha
=\{a_{1},a_{2},a_{3},a_{4},a_{1}^{\dagger },a_{2}^{\dagger },a_{3}^{\dagger
},a_{4}^{\dagger }\}$, which satisfies $\alpha ^{\dagger }=T\beta ^{\dagger }
$ with a transformation matrix $T$. To ensure the bosonic commutation
relations for the operators, the matrix $T$ satisfies
\begin{equation}
T^{\dagger }\Lambda T=\Lambda =\left(
\begin{array}{cc}
I_{4} & 0 \\
0 & -I_{4}%
\end{array}%
\right) ,  \label{Tmatrix}
\end{equation}%
where $I_{4}$ is the identity matrix with the dimension $4\times 4$.
Substituting for $\alpha $ and $\alpha ^{\dagger }$ in terms of $\beta
^{\dagger }$ and $\beta $, one obtains $H_{\text{cCP}}^{\downarrow }=\alpha
M\alpha ^{\dagger }=\beta T^{\dagger }MT\beta ^{\dagger }$ in diagonalized
form as
\begin{equation}
H_{\text{cCP}}^{\downarrow }=\beta \varepsilon \beta ^{\dagger
}=2\sum_{n=1}^{4}\varepsilon _{n}b_{n}^{\dagger }b_{n}+\frac{1}{2}%
\varepsilon _{n},
\end{equation}%
where the matrix $\varepsilon $ of the eigenvalues $\{\varepsilon _{i}\}$ is
\begin{equation}
\varepsilon =T^{\dagger }MT=\left(
\begin{array}{cccccccc}
\varepsilon _{1} & 0 & 0 & 0 & 0 & 0 & 0 & 0 \\
0 & \varepsilon _{2} & 0 & 0 & 0 & 0 & 0 & 0 \\
0 & 0 & \varepsilon _{3} & 0 & 0 & 0 & 0 & 0 \\
0 & 0 & 0 & \varepsilon _{4} & 0 & 0 & 0 & 0 \\
0 & 0 & 0 & 0 & \varepsilon _{1} & 0 & 0 & 0 \\
0 & 0 & 0 & 0 & 0 & \varepsilon _{2} & 0 & 0 \\
0 & 0 & 0 & 0 & 0 & 0 & \varepsilon _{3} & 0 \\
0 & 0 & 0 & 0 & 0 & 0 & 0 & \varepsilon _{4}%
\end{array}%
\right) .
\end{equation}%
It follows that $T^{\dagger }MT=\Lambda T^{-1}\Lambda MT=\varepsilon $,
resulting in $T^{-1}\Lambda MT=\Lambda \varepsilon $. Hence, the eigenvalues $\pm
\varepsilon _{i}$ are obtained by diagonalizing the matrix $\Lambda M$ with
the matrix $M$ in Eq.(\ref{Mmatrix}). Similarly, one can diagonalize $%
\Lambda M$ to obtain the excitation energy for $N=3$.

\color{black}
\section{Numerical mean-field phase diagram of the LMG ring for N=3 and N=4}

\begin{figure}[ht!]
\includegraphics[scale=0.4]{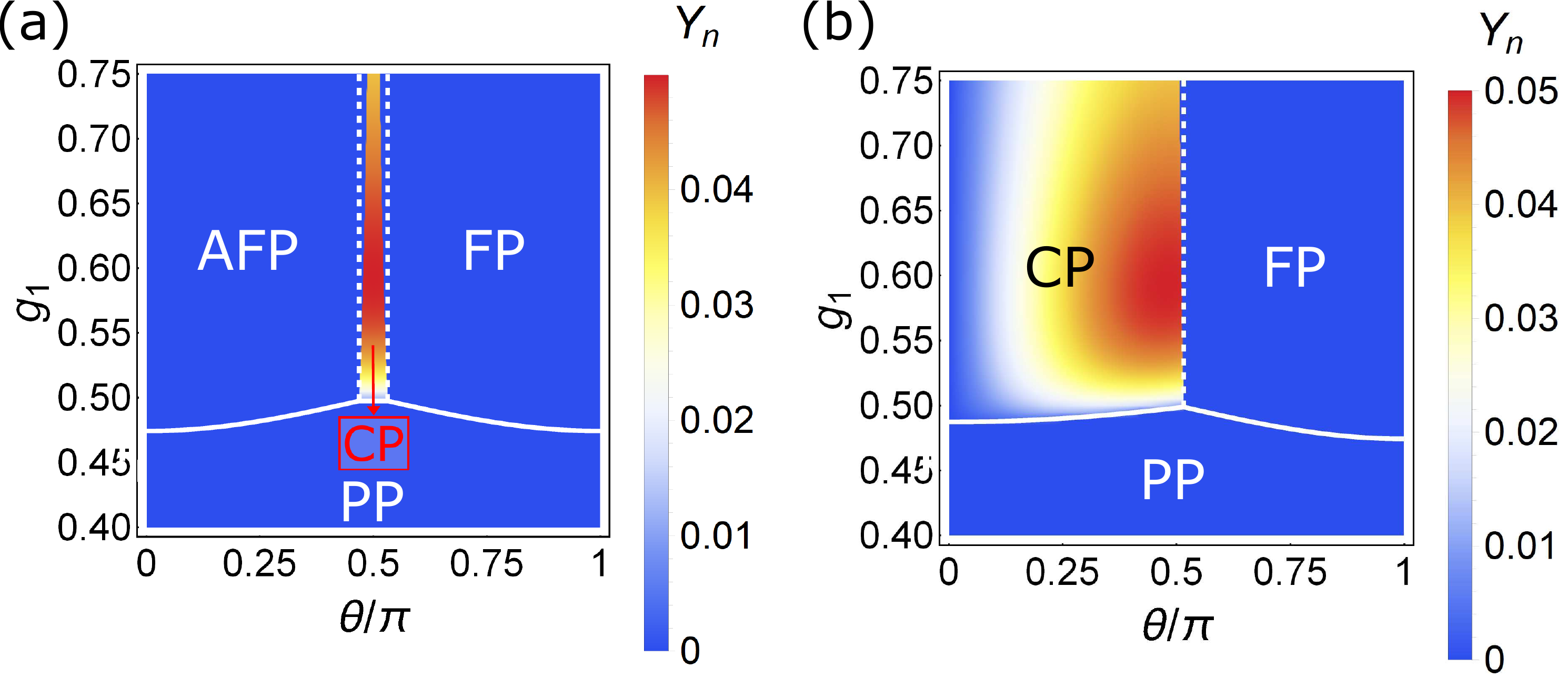}
\caption{Phase diagram in the $\theta-g_1$ plane for the LMG ring using $Y_n$ for a given site $n$ as order parameter for (a) $N=4$ and (b) $N=3$. To facilitate visualization of the different phases we have chosen one of the degenerate configurations of the ground state for each phase, such that $Y_n>0$ for the chiral phase. The solid white line represents the second-order phase boundary, while vertical dashed lines represent the predicted first-order lines obtained for the equivalent quantum Rabi ring. The hopping strength is set to $J/\omega =0.05$.}
\label{PDSUPP1}
\end{figure}

It's important to note that an effective Hamiltonian for the non-paramagnetic phases of the model on Eq.~\ref{SLMGR} will, in general, not have the form of Eq.~\ref{SEFFSR}. However, as shown in Fig.\ref{PDSUPP1} and Figs.(1) and (2) of the main text, the mean-field order parameter values $X_n$ and $Y_n$ found through minimization of Eq.\ref{SMFLMGR} are in very good agreement with the phase boundaries found for the quantum Rabi Ring. For both $N=4$ and $N=3$ we can observe that the second-order boundaries fit exactly the numerical values of $X_n$, this is not surprising as we already discussed that both systems share the same second-order boundaries.

It's also helpful to study the behavior of $Y_n$ as a non-zero value of this parameter differentiates the chiral magnetic phase from the other non-paramagnetic phases. We see that as we move deeper in the non-paramagnetic phases (increasing $g_1$) the numerical first order boundary starts to deviate from the one predicted in the quantum Rabi ring (vertical dashed line), however, the results are almost identical for a very good portion of the phase diagram. Making the quantum Rabi ring an exciting candidate to study chiral magnetism.

\subsection{Analytical expressions of the order parameter for the quantum Rabi square}
In this section, we derive the analytical expression of phase boudaries $g_{1c}^{F}$, $g_{1c}^{AF}$, $g_{1c}^{C}$  in the
ferromagnetic superradiant phase (FSP),
anti-ferromagnetic superradiant phase (AFSP) and chiral superradiant phase (CSP), and the critical value $\theta_c^{\pm}$ given in Eq.($7$) of the main text, respectively. The analytical solutions are given in
detail in the infinite frequency limit $\Delta/\omega\rightarrow\infty$.

It is required that the off-diagonal term $V_{off}$ in Eq.(\ref{offdiagonal}) vanishes, which gives
\begin{equation}
\omega \alpha _{n}-g\sin (2\gamma _{n})+J(e^{i\theta }\alpha
_{n+1}+e^{-i\theta }\alpha _{n-1})=0,
\end{equation}%
One obtains the equations for the real part ${Re}(V_{off})=0$ and for the
imaginary part ${Im}(V_{off})=0$%
\begin{equation}\label{real1}
\omega A_{n}-\frac{4g^{2}A_{n}}{\sqrt{16g^{2}A_{n}^{2}+\Delta ^{2}}}+J\cos
\theta (A_{n+1}+A_{n-1})+J\sin \theta (B_{n-1}-B_{n+1})=0,  
\end{equation}
\begin{equation}
\omega B_{n}+J\sin \theta (A_{n+1}-A_{n-1})+J\cos \theta (B_{n+1}+B_{n-1})=0.
\end{equation}%
It is equivalent to the equations obtained by minimizing energy in Eq.\ref{SMFQRR} with
respect to all $A_{n}$'s and $B_{n}$'s.

It is easy to obtain $\sum_{n}{Im}(V_{off})=0$, giving $\omega
\sum_{n}B_{n}+2J\cos \theta \sum_{n}B_{n}=0$. This leads to $\sum_{n}B_{n}=0$
and $B_{n}=-(B_{n+1}+B_{n+2}+B_{n-1})$, and to $\sum_{n}B_{n}=0$.
Using the periodic boundary condition $B_{N+1}=B_{1}$ and $A_{N+1}=A_{1}$, we
obtain the conditions for the even and odd sites as%
\begin{equation}
\sum_{k=0}B_{2k+1}=0,\sum_{k=1}B_{2k}=0.
\end{equation}
In particular, when $N$ is odd, there are additional conditions $B_{N}=-B_{1}$ and $A_{N}=A_{1}$.

Consequently, we obtain the constrain for the imaginary part of
$\alpha _{n}$ for $N=4$
\begin{equation}\label{imag}
B_{n+1}=-B_{n-1},B_{n}=-\frac{J\sin \theta }{\omega }(A_{n+1}-A_{n-1}).
\end{equation}

By substituting $B_{n}$ into the Eq.(\ref{real1}), one obtains
\begin{equation}\label{real}
A_{n}(\omega -\frac{4g^{2}}{\sqrt{16g^{2}A_{n}^{2}+\Delta ^{2}}}-\frac{%
2J^{2}\sin ^{2}\theta }{\omega })+J\cos \theta (A_{n+1}+A_{n-1})+\frac{%
2J^{2}\sin ^{2}\theta }{\omega }A_{n+2}=0.
\end{equation}%
The analytical solutions of $A_{n}$ and $B_{n}$ are derived below.

\begin{figure}[t]
\includegraphics[scale=0.3]{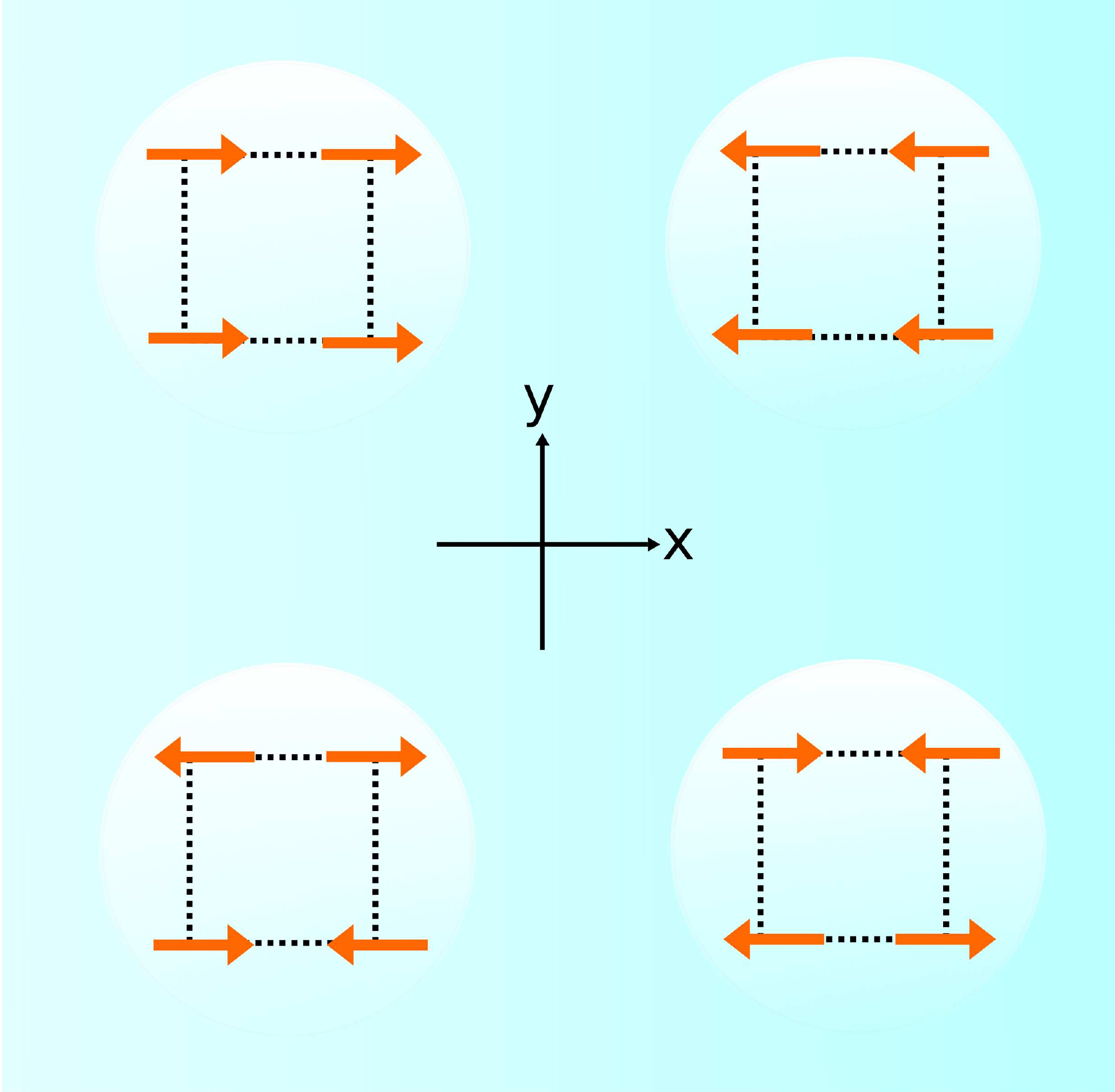}
\caption{Allowed configurations in the ferromagnetic phase (top row) and antiferromagnetic phase (bottom row). Arrows represent the magnetization on the x-y plane on each site. A similar representation can be done for the ferro-superradiant phase (top) and the antiferro-superradiant phase (bottom) if $A_n$ and $B_n$ are represented in the complex plane.}
\label{ConfSupp1}
\end{figure}

\subsubsection{(i)Ferro/Antiferro-superradiant phase
(FSP,AFSP) }
In the FSP and AFSP, all order parameters are real. It implies that $B_{n}=0$. According to constraint conditions of $B_n$ in Eq. (\ref{imag}), one has $%
A_{1}=A_{3}=a,A_{2}=A_{4}=a'$. Substituting them into Eq. (\ref{real}) of $A_n$ leads to
\begin{eqnarray}
a(\omega -\frac{4g^{2}}{\sqrt{16g^{2}a^{2}+\Delta ^{2}}}-\frac{2J^{2}\sin
^{2}\theta }{\omega })+2Ja'\cos \theta +\frac{2J^{2}\sin ^{2}\theta }{\omega }%
a &=&0, \\
a'(\omega -\frac{4g^{2}}{\sqrt{16g^{2}a^{'2}+\Delta ^{2}}}-\frac{2J^{2}\sin
^{2}\theta }{\omega })+2Ja\cos \theta +\frac{2J^{2}\sin ^{2}\theta }{\omega }%
a' &=&0.
\end{eqnarray}%
One solution is
\begin{equation}
a=a'=\pm \frac{1}{4g}\sqrt{\frac{16g^{4}}{(\omega +2J\cos \theta )^{2}}%
-\Delta ^{2}}.
\end{equation}%
Setting this equation to zero, one finds the critical strength as $g_{1c}^{\texttt{FSP}}=\frac{1}{2}\sqrt{1+2J/\omega
\cos \theta }$, which is consistent with setting $q=0$ in Eq.\ref{Scriticalg}. Then $A_{n}$ is the same
for all four cavities, and hence we call the phase ferro-superradiant phase due to its similarities with a ferromagnetic phase (see Table 1 of the main text).

The other solution is
\begin{equation}
a=-a'=\pm \frac{1}{4g}\sqrt{\frac{16g^{4}}{(\omega -2J\cos \theta )^{2}}%
-\Delta ^{2}},
\end{equation}%
Which leads to the critical coupling strength $g_{1c}^{\texttt{AFSP}}=\frac{1}{2}\sqrt{
1-2J/\omega \cos \theta }$, consistent with setting $q=\pi$ in Eq.\ref{Scriticalg}. In this phase $A_1=A_3=-A_2=-A_4$, which mimics an antiferromagnetic order, hence, we call the phase antiferro-superradiant phase. 

Fig.~\ref{ConfSupp1} shows a schematic representation of the allowed degenerate configurations in the AFSP and FSP.

\subsubsection{(ii)Chiral superradiant phase (CSP)}

In the CSP, $\alpha_{n}$ is complex and depends on $n$.
Considering $B_{n}\neq0$, we have $A_{1}\neq A_{3}$ and $%
A_{2}\neq A_{4}$. It includes two kinds of solutions $%
A_{1}=A_{4}=a,A_{2}=A_{3}=a'$ and $A_{1}=A_{2}=a,A_{3}=A_{4}=a'$. For the first case $%
A_{1}=A_{4}=a, A_{2}=A_{3}=a'$, equations of $A_{n}$ reduce to
\begin{eqnarray}
a(\omega -\frac{4g^{2}}{\sqrt{16g^{2}a^{2}+\Delta ^{2}}}-\frac{2J^{2}\sin
^{2}\theta }{\omega })+J\cos \theta (a+a')+\frac{2J^{2}\sin ^{2}\theta }{%
\omega }a' &=&0, \\
a'(\omega -\frac{4g^{2}}{\sqrt{16g^{2}a'^{2}+\Delta ^{2}}}-\frac{2J^{2}\sin
^{2}\theta }{\omega })+J\cos \theta (a'+a)+\frac{2J^{2}\sin ^{2}\theta }{%
\omega }a &=&0.
\end{eqnarray}
One solution is
\begin{equation}
a=-a'=\pm \frac{1}{4g}\sqrt{\frac{16\omega ^{2}g^{4}}{(\omega ^{2}-4J^{2}\sin
^{2}\theta )^{2}}-\Delta ^{2}.}
\end{equation}%
It leads to the critical coupling strength $g_{1c}^{\texttt{CSP}}=\frac{1}{2}\sqrt{%
1-4J^{2}/\omega ^{2}\sin ^{2}\theta }$, consistent with setting $q=\pm \pi/2$ in Eq.\ref{Scriticalg}. The values of the imaginary parts
are
\begin{equation*}
B_{1}=\frac{2J\sin \theta }{\omega }a'=B_{2}.
\end{equation*}%
It is equivalent to the solution of the case with $A_{1}=A_{4}=a',A_{2}=A_{3}=a$.

On the other hand, for the second case $A_{1}=A_{2}=a,A_{3}=A_{4}=a'$,
one obtains%
\begin{eqnarray}
a(\omega -\frac{4g^{2}}{\sqrt{16g^{2}a^{2}+\Delta ^{2}}}-\frac{2J^{2}\sin
^{2}\theta }{\omega })+J\cos \theta (a'+b)+\frac{2J^{2}\sin ^{2}\theta }{%
\omega }a' &=&0, \\
a'(\omega -\frac{4g^{2}}{\sqrt{16g^{2}a^{'2}+\Delta ^{2}}}-\frac{2J^{2}\sin
^{2}\theta }{\omega })+J\cos \theta (a'+a)+\frac{2J^{2}\sin ^{2}\theta }{%
\omega }a &=&0.
\end{eqnarray}
The solutions of $a$ and $a'$ are the same as the above results of the first
case. But the values of imaginary parts are different
\begin{equation}
B_{1}=-\frac{2J\sin \theta }{\omega }a'=-B_{2}.
\end{equation}

Fig.~2 of the main text shows the four allowed configurations in the CSP.

\end{document}